\documentclass[aps,prb,reprint,groupedaddress,superscriptaddress,showpacs]{revtex4-1}

\usepackage{amsmath}

\usepackage{graphics}

\usepackage{graphicx}

\usepackage{epsfig}

\usepackage{amsmath}

\usepackage{amsfonts}

\usepackage{amssymb}%

\usepackage{setspace}   %Allows double spacing with the \doublespacing command

\usepackage{xcolor}

\providecommand{\U}[1]{\protect\rule{.1in}{.1in}}

\begin{document}

\title{Converse effect of pressure on quadrupolar and magnetic transition in Ce$_{3}$Pd$_{20}$Si$_{6}$}

\author{J. \surname{Larrea J.}}

\email[E-mail me at: ]{jlarreaj@cbpf.br}

\affiliation{Institute of Solid State Physics, Vienna University of Technology, Wiedner Haupst. 8~-~10, 1040 Vienna, Austria.}

\affiliation{Department of Physics, University of Johannesburg, Auckland Park 2006, South Africa.}

\affiliation{Centro Brasileiro de Pesquisas F\'isicas, Rua Doutor Xavier Sigaud 150, CEP 22290-180, Brazil.}

\altaffiliation[Current address: ]{Centro Brasileiro de Pesquisas F\'isicas, Rua Doutor Xavier Sigaud 150, CEP 22290-180, Brazil.}

\author{A. M. \surname{Strydom}}

\affiliation{Department of Physics, University of Johannesburg, Auckland Park 2006, South Africa.}

\author{V. \surname{Martelli}}

\affiliation{Institute of Solid State Physics, Vienna University of Technology, Wiedner Haupst. 8~-~10, 1040 Vienna, Austria.}

\affiliation{Centro Brasileiro de Pesquisas F\'isicas, Rua Doutor Xavier Sigaud 150, CEP 22290-180, Brazil.}

\author{A. \surname{Prokofiev}}

\author{K.-A \surname{Lorenzer}}

\affiliation{Institute of Solid State Physics, Vienna University of Technology, Wiedner Haupst. 8~-~10, 1040 Vienna, Austria.}

\author{H. M. \surname{R${\o}$nnow}}

\affiliation{Laboratory for Quantum Magnetism, \'Ecole Polytechnique F\'ed\'erale de Lausanne, 1015 Lausanne, Switzerland.}

\author{S. \surname{Paschen}}

\affiliation{Institute of Solid State Physics, Vienna University of Technology, Wiedner Haupst. 8~-~10, 1040 Vienna, Austria.}

\date{\today}

\begin{abstract}

The heavy fermion compound Ce$_{3}$Pd$_{20}$Si$_{6}$ displays unconventional quantum criticality as the lower of two consecutive phase transitions is fully suppressed by magnetic field. Here we report on the effects of pressure as additional tuning parameter. Specific heat and electrical resistivity measurements reveal a converse effect of pressure on the two transitions, leading to the merging of both transitions at 6.2 kbar. The field-induced quantum criticality is robust under pressure tuning. We rationalize our findings within an extended version of the global phase diagram for antiferromagnetic heavy fermion quantum criticality.

%\textcolor{red}{quantum critical point}

\end{abstract}


\maketitle

\section{Introduction}\label{}

%\doublespacing

Quantum criticality in heavy fermion systems continues to attract great attention [\onlinecite{lohn, Gegenwart, si2010, si2013}]. The ground state of these materials is determined by the competition between the Ruderman$-$Kittel$-$Kasuya$-$Yosida (RKKY) interaction and the Kondo interaction. When the RKKY interaction dominates a magnetic, typically antiferromagnetic, ground state is realized. The opposite case results in a paramagnetic heavy fermion state. By applying a non-thermal control parameter such as pressure or magnetic field, transitions between the two phases can frequently be realized. If the suppression of the finite-temperature phase transition remains continuous a quantum critical point (QCP) is accessed.

Various scenarios have been proposed to describe quantum critical behavior. The spin-density wave scenario [\onlinecite{Hertz, Mills, Moriya}] attributes all effects to the suppression of the order parameter and the critical fluctuations associated with it. Other scenarios assume that a second mode is critical at the QCP. In the theory of local quantum criticality [\onlinecite{Si}] this is the Kondo interaction. This Kondo breakdown scenario was argued to describe the quantum criticality of various heavy fermion (HF) compounds [\onlinecite{lohn},\onlinecite{Gegenwart},\onlinecite{si2013}] much better than the spin-density wave scenario. The Kondo breakdown at the border between an antiferromagnetic (AF) and paramagnetic (PM) state requires the presence of quasi-two dimensional spin fluctuations. These are not unlikely to be present in systems such as YbRh$_{2}$Si$_{2}$ [\onlinecite{Trovarelli},\onlinecite{custers2003}], CeRhIn$_{5}$ [\onlinecite{hegger}] or CeCu$_{6-x}$Au$_{x}$ [\onlinecite{lohn2}], which show strong magnetic anisotropy. More recently, signatures of Kondo breakdown were observed in the magnetic-field induced QCP of the cubic system Ce$_{3}$Pd$_{20}$Si$_{6}$ [\onlinecite{cus2012}]. Since this system is isotropic at the QCP, this raised questions about the role of dimensionality in Kondo breakdown quantum criticality. One way to reconcile the experimental observation with the suggested global phase diagram for AF heavy fermion quantum criticality [\onlinecite{si2010}] is to assume that field-induced magnetic order is present within the putative antiferroquadrupolar phase below $T_{Q}(B)$ [\onlinecite{si2013},\onlinecite{cus2012}]. In this case, the Kondo breakdown transition could be seen as a small Fermi surface to large Fermi surface transition within the AF portion of the global phase diagram at low values of the frustration parameter $G$ that corresponds to the 3D limit [\onlinecite{cus2012}].

The cubic HF compound Ce$_{3}$Pd$_{20}$Si$_{6}$ crystallizes in a Cr$_{23}$C$_{6}$-type structure with space group $Fm\overline{3}m$ [\onlinecite{grib1994}]. The Ce atoms in Ce$_{3}$Pd$_{20}$Si$_{6}$ occupy two sites with different cubic point symmetry. At the 4$a$ site ($O_h$ symmetry) the Ce atoms are positioned inside a cage of 12 Pd atoms and 6 Si atoms whereas at the 8$c$ site ($T_d$ symmetry) the Ce atoms are surrounded by 16 Pd atoms. In polycrystalline samples, two successive phase transitions are observed at $T_N$ = 0.3 K and $T_Q$ = 0.5 K and have tentatively been attributed to AF and to antiferroquadrupolar (AFQ) order, respectively [\onlinecite{strydom2006}]. The crystal electric field (CEF) scheme at the two Ce sites is still a matter of debate [\onlinecite{goto2009,mitamura2010, deen2010}]. The suppression of $T_N$ to zero at $B_C$ $\approx$ 0.9 T leads to a field-induced QCP with Kondo breakdown [\onlinecite{cus2012}]. In polycrystalline samples, signatures of $T_Q$ can be discerned in magnetic fields up to at least 10 T (Ref. [\onlinecite{cus2012}]). Recent investigations on single crystals under magnetic field ($B$) revealed that at fields above 1 T, $T_Q(B)$ is anisotropic with respect to the direction along which $B$ is applied [\onlinecite{goto2009},\onlinecite{mitamura2010},\onlinecite{ono},\onlinecite{martelli}].

An alternative route to quantum criticality in Ce$_{3}$Pd$_{20}$Si$_{6}$ might be to use pressure as control parameter. The critical pressure necessary to fully suppress $T_N$ was estimated to be 5 kbar [\onlinecite{prok2009}]. Electrical resistivity and specific heat investigations on a lower quality polycrystalline Ce$_{3}$Pd$_{20}$Si$_{6}$ sample up to 80 kbar (8 GPa) in temperatures down to 0.5 K revealed an increase of the Kondo temperature ($T_K$) with pressure [\onlinecite{hashiguchi}]. However, no information about the pressure evolution of $T_N$ or $T_Q$ could be inferred from those measurements. More recently, electrical resistivity measurements up to 40 kbar in the isostructural germanide compound Ce$_{3}$Pd$_{20}$Ge$_{6}$ revealed that both the AF ($T_N$ = 0.75 K) and the ferroquadrupolar ($T_{FQ}$ = 1.2 K) transition show a tendency to disappear at pressures higher than 50 kbar [\onlinecite{hidaka}].

Here, we present a study of the pressure$-$magnetic field$-$temperature phase diagram of Ce$_{3}$Pd$_{20}$Si$_{6}$ using hydrostatic pressure conditions. Our aim is to investigate how $T_N$ and $T_Q$ evolve with pressure in the range where the pressure-tuned AF QCP was predicted [\onlinecite{prok2009}]. We also explore whether the field $-$induced quantum criticality is modified under pressure.

\section{Experimental}\label{}

Polycrystalline Ce$_{3}$Pd$_{20}$Si$_{6}$ samples were synthesized by melting Ce, Pd and Si in a horizontal water-cooled copper boat using high-frequency heating. Details on the synthesis and characterization are described elsewhere [\onlinecite{prok2009}]. The sample used for the present study is of the same quality as those reported in previous works [\onlinecite{cus2012},\onlinecite{strydom2006},\onlinecite{prok2009}]. Electrical resistivity and specific heat measurements were performed in a CuBe piston-cylinder pressure cell for pressure up to 6.2 kbar, with kerosene as pressure transmitting medium and Pb as in-situ manometer. All electrical contacts were spot welded onto the same piece of sample with dimensions 3.0 mm$\times$2.0 mm$\times$0.25 mm. The conventional four-probe AC method was used to measure electrical resistivity. Specific heat was measured by AC calorimetry. For the latter, a constantan wire and a pair of Au-Fe(0.07$\%$) and chromel wires of 25 $\mu$m diameter were used as a heater and thermocouple, respectively. An oscillating excitation current with $\omega =$ 0.5 Hz and $I$ = 0.2 mA was applied to the sample heater. The sample modulation temperature was read out by a lock-in amplifier in a second harmonic mode and recorded as pick-up voltage ($V_{ac}$). The inverse of this quantity is approximately proportional to the sample's specific heat [\onlinecite{larrea1}]. The sample temperature was corrected for a DC offset due to Joule heating by separately measuring the temperature signal with a DC nanovoltmeter. This correction was found to be at maximum 0.02 K at the lowest temperature. The pressure cell was inserted into a $^{3}$He/$^{4}$He dilution refrigerator with a superconducting magnet to measure both $\rho (T)$  and $C_p(T)$ down to 0.05 K and under magnetic field up to $B = \mu_0 H$ $=$ 14 T. The magnetoresistance measurements at constant temperature and pressure were carried out with a field sweep of 50 mT/min and with a temperature stabilization of $\pm$ 1 mK.

\section{Results and discussion}\label{}

\subsection{Electrical transport}\label{}

\begin{figure}[htbp]

\centering \includegraphics[angle=0,scale=0.4]{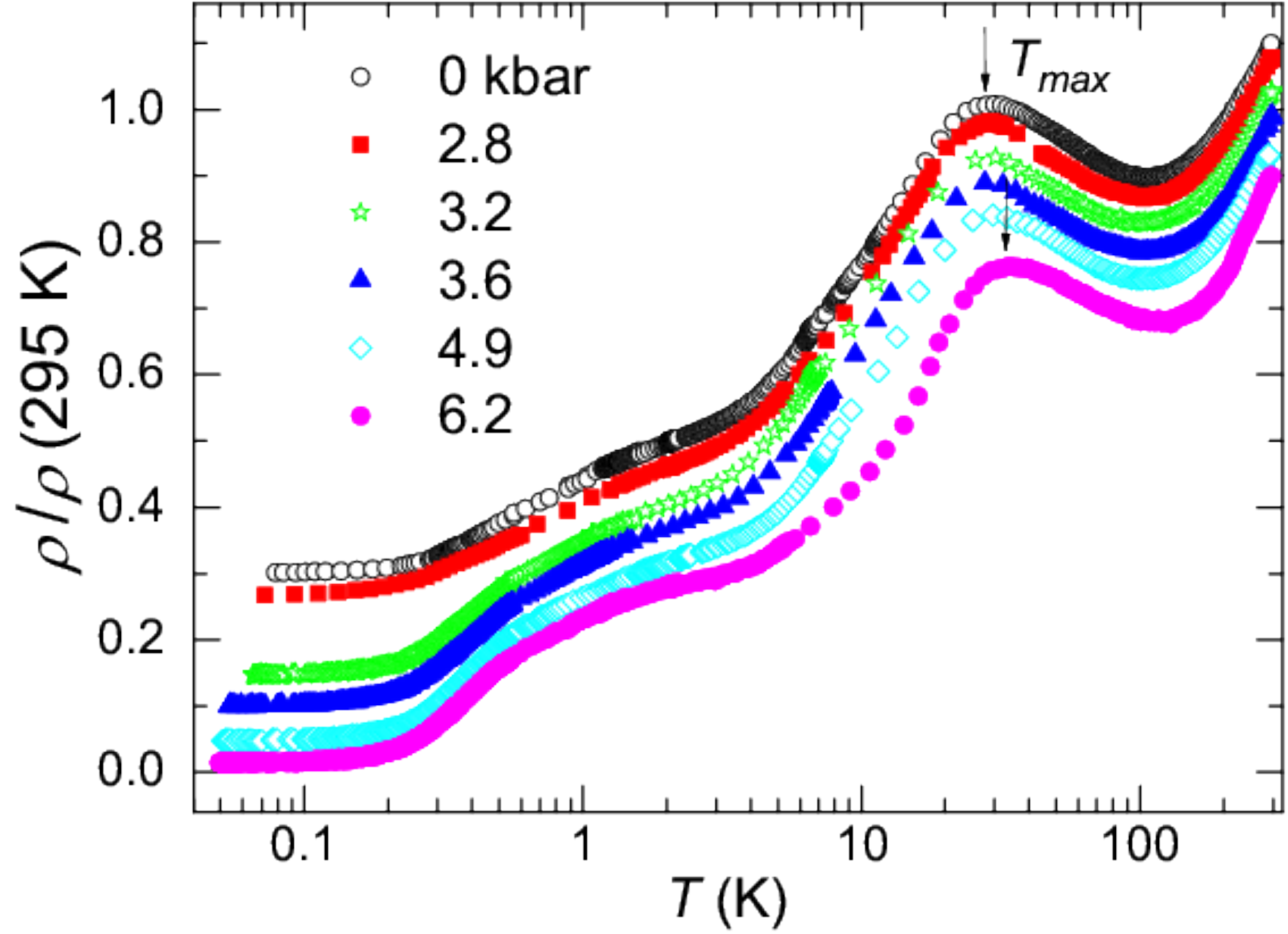}

\caption{(Color online) Temperature dependence of the electrical resistivity normalized to the room temperature value ($\rho / \rho_{295K}$) at different pressures. For better readability the data are shifted by +0.1 at 0 kbar, +0.075 at 2.8 kbar, +0.025 at 3.2 kbar, -0.025 at 3.6 kbar, -0.075 at 4.9 kbar, -0.1 at 6.2 kbar. The arrows indicate $T_{max}$, the temperature where $d\rho/dT$ is zero.  \label{fig1}}

\end{figure}

The temperature dependence of the electrical resistivity ($\rho (T)$), normalized to the room temperature value ($\rho_{295K}$) at different pressures ($p$), is shown in Fig.~\ref{fig1}. For clarity data are shifted by fixed amounts (see caption). With decreasing temperature, for all pressures, $\rho$ first decreases until it reaches a local minimum around $\sim$ 120 K, then increases roughly as -ln$T$ until it develops a broad maximum around $T_{max}$. Below $T_{max}$, $\rho(T)$ falls rapidly, showing an $S$-shaped profile below 3 K, with a broad hump around 1 K. This profile is similar to $\rho(T)$ data previously reported for $p <$ 80 kbar and $T >$ 0.5 K in a sample that showed a lower $T_N$ and no clear sign of quadrupolar order [\onlinecite{hashiguchi}].

The maximum and the kink might either be due to Kondo scattering from the excited and ground state crystal electric field levels, respectively, as expected in the Cornut and Coqblin scenario [\onlinecite{coublin}], or due to Kondo scattering from the two different Ce sites of the crystal structure as suggested previously [\onlinecite{hashiguchi},\onlinecite{hidaka}]. In either case, we expect the temperature of the maximum ($T_{max}$) to contain information on the Kondo temperature of at least one site. We determine $T_{max}$ in Fig.~\ref{fig1} as the temperature where $d\rho(T)/dT$ is zero. $T_{max}$ increases slightly with pressure, in particular above 4 kbar (Fig.~\ref{fig4}a). A positive slope $dT_{max}/dP$ was also reported at higher pressures [\onlinecite{hashiguchi}]. Assuming that the relatively low pressure does not sizably affect the CEF level scheme, the increase of $T_{max}$ with $p$ may be associated with an increase of the Kondo interaction due to an enhancement of the effective hybridization between the Ce 4$f$ and the conduction electrons.

\begin{figure}[htbp]

\centering \includegraphics[angle=0,scale=0.4]{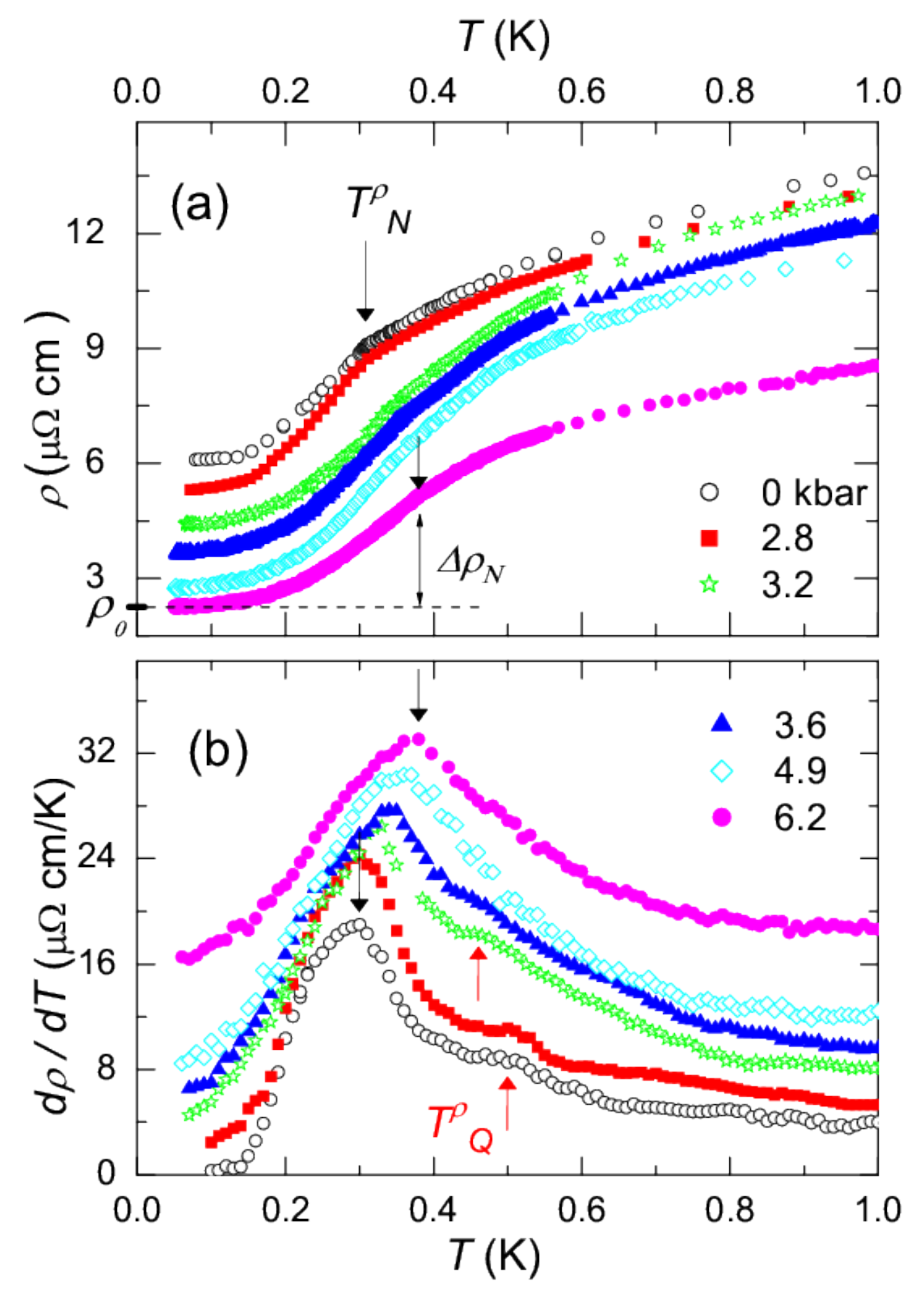}

\caption{(Color online) (a) Electrical resistivity ($\rho$) below 1 K at different pressures. $\rho(T)$ data are shifted by -0.5 at 2.8 kbar, -0.75 at 3.2 kbar, -1.4 at 3.6 kbar, -2.2 at 4.9 kbar and -1 at 6.2 kbar ($\mu\Omega$cm). $\rho_{0}$ is the residual resistivity at $T =$ 0 and $\Delta\rho_{N}$ the resistivity change up to $T^{\rho}_{N}$. (b) Temperature dependence of the first derivative of the electrical resistivity ($d\rho/dT$). The curves were subsequently shifted by +2$\mu\Omega$cm/K and at 6.2 kbar by +16$\mu\Omega$cm/K. The downward and upward arrows indicate the temperature where $d\rho/dT$ shows a maximum ($T^{\rho}_{N}$) and shoulder-type feature ($T^{\rho}_{Q}$), which are ascribed tentatively to the onset of antiferromagnetic and antiferroquadrupolar order, respectively. \label{fig2}}

\end{figure}

\begin{figure}[htbp]

\centering \includegraphics[angle=0,scale=0.45]{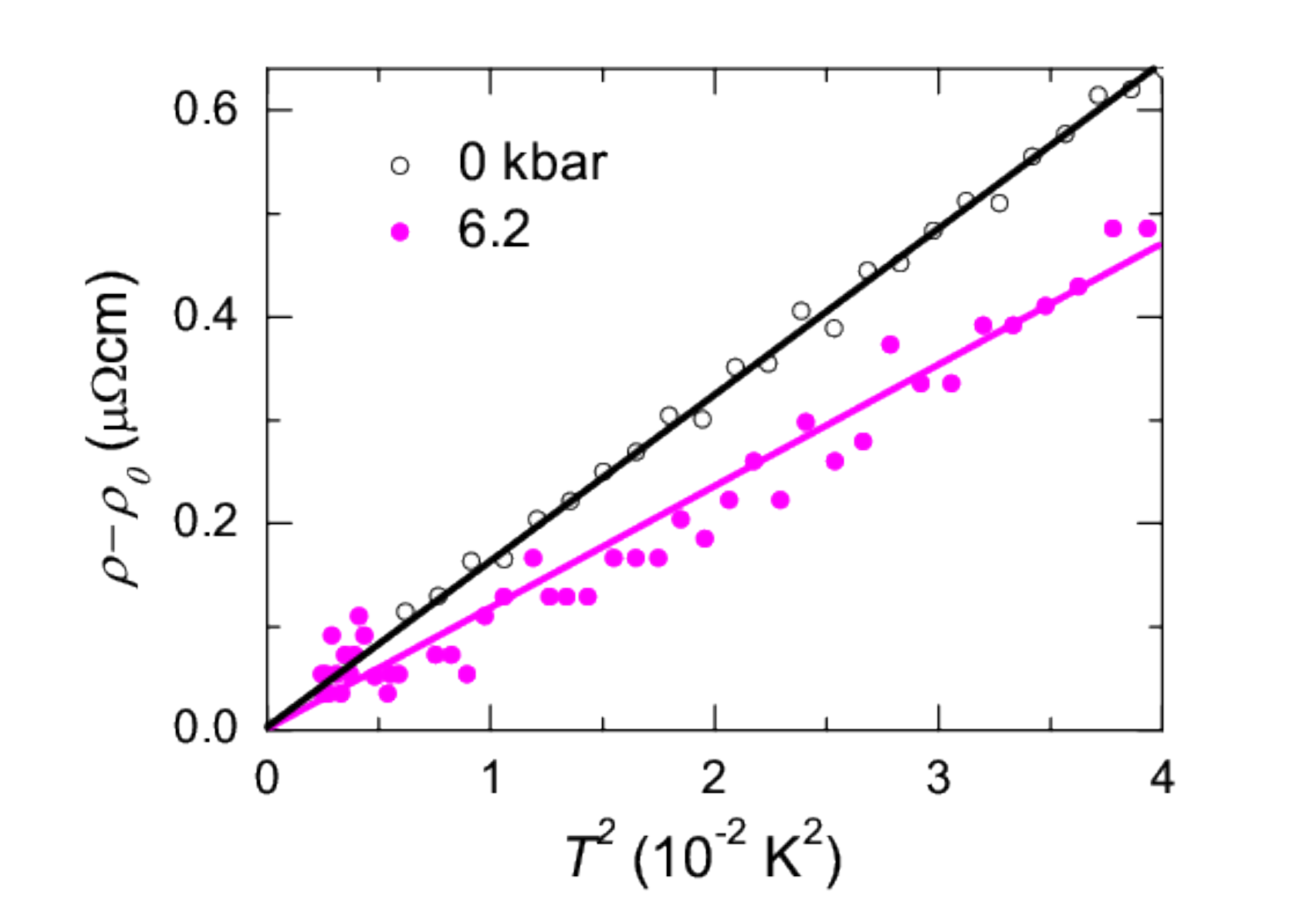}

\caption{(Color online) Difference of electrical resistivity and residual resistivity plotted versus $T^{2}$ for 0 kbar (open symbols) and 6.2 kbar (full symbol). The lines are linear fits (see text). \label{fig3}}

\end{figure}

\begin{figure}[htbp]

\centering \includegraphics[angle=0,scale=0.4]{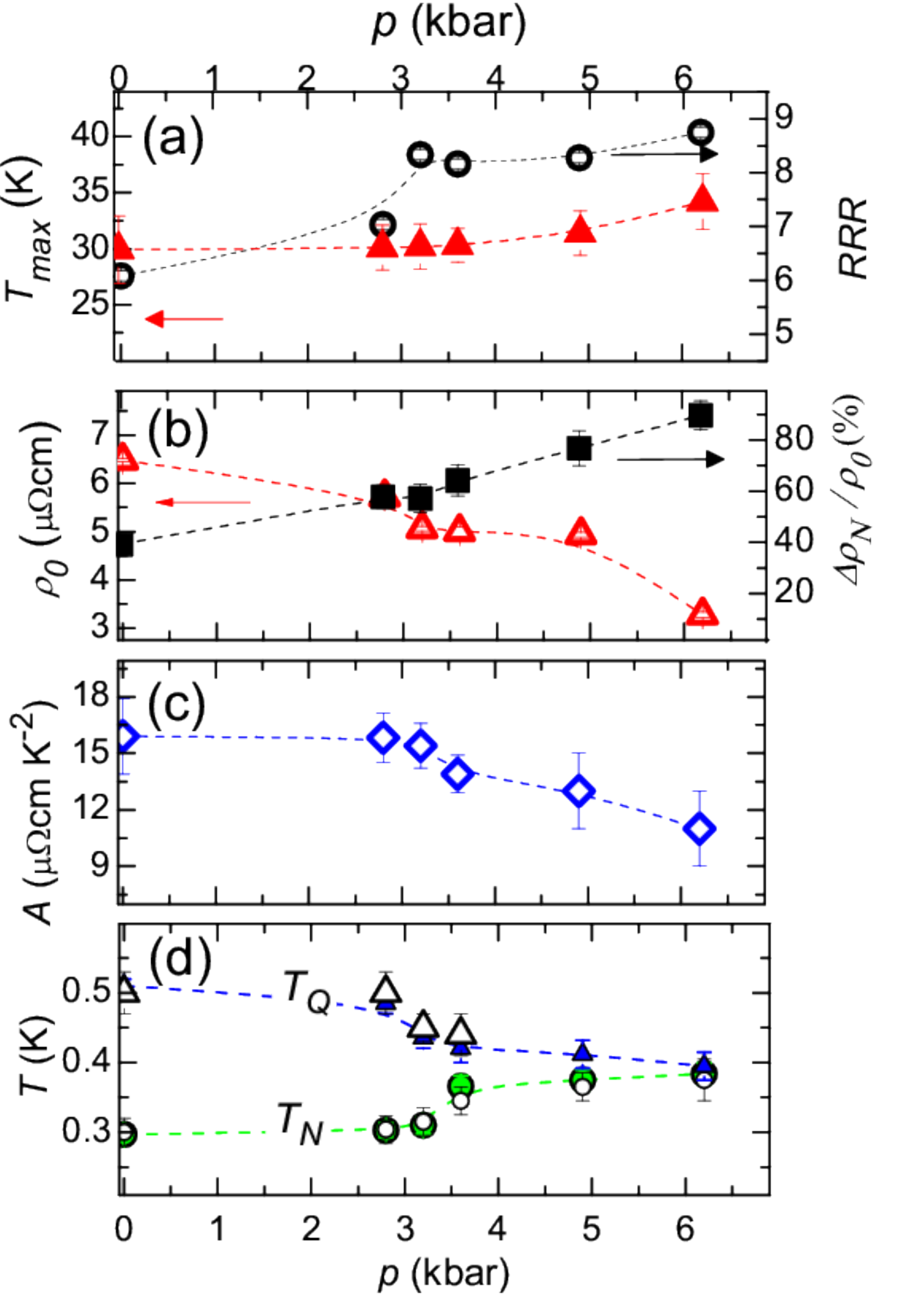}

\caption{(Color online) (a) Pressure dependence of $T_{max}$ of Fig.~\ref{fig1} and the residual resistance ratio (RRR) (see text). (b) Pressure variation of the residual resistivity ($\rho_{0}$) determined by fitting $\rho=\rho_{0} + AT^{2}$ to the date below 0.2 K and of $\Delta\rho_{N}$/$\rho_{0}$, the relative change of $\rho(T)$ up to $T^{\rho}_{N}$ estimated as shown in Fig.~\ref{fig2}a. (c) Fermi liquid $A$ coefficient as a function of pressure obtained from fitting as in (b). (d) Temperature-pressure phase diagram, where $T_{N}$ and $T_Q$ are estimated from electrical resistivity (open symbols, $T^{\rho}_{N}$ and $T^{\rho}_{Q}$) and $C_p$ data (full symbols). Dashed lines are guides to the eyes. \label{fig4}}

\end{figure}

The electrical resistivity at different pressures at the lowest temperatures is shown in Fig.~\ref{fig2}a. For all pressures, $\rho(T)$ first decreases gradually with decreasing temperature down to $\sim$ 0.6 K, then more steeply below 0.6 K, and finally tends to saturate below $\sim$ 0.15 K. This behavior is typical of antiferromagnetic heavy fermion metals where the magnetic ordering temperature is associated either with the position of a kink in $\rho(T)$ or with the temperature where the first derivative of the electrical resistivity ($d \rho/dT$) shows a maximum [\onlinecite{lohn}]. Here, we use the latter criterion to determine $T^{\rho}_{N}$, as is shown in Fig.~\ref{fig2}b. At $p =$ 0 kbar, we can distinguish a clear maximum at $T^{\rho}_{N}=$ 0.3 K and a shoulder at $T^{\rho}_{Q}$ = 0.5 K. $T^{\rho}_{N}$ and $T^{\rho}_{Q}$ are tentatively assigned to the onset of antiferromagnetic and antiferroquadrupolar order, respectively. With increasing pressure up to 3.6 kbar, these two features follow converse trends, i.e., $T^{\rho}_{N}$ is enhanced whereas $T^{\rho}_{Q}$ is reduced (Fig.~\ref{fig4}d, open symbols). For pressures above 4.9 kbar, $d\rho/dT$ broadens. This is likely due to the fact $T^{\rho}_{N}$ and $T^{\rho}_{Q}$ are too close to be distinguished. $T_N(p)$ increases in the whole investigated pressure range, with a small step-like feature at 3.2 kbar. At the same pressure a pronounced increase of the residual resistance ratio RRR $= \rho_{295K}/\rho_{0.05K}$ is observed (Fig.~\ref{fig4}a).

\begin{figure}[htbp]

\centering \includegraphics[angle=0,scale=0.45]{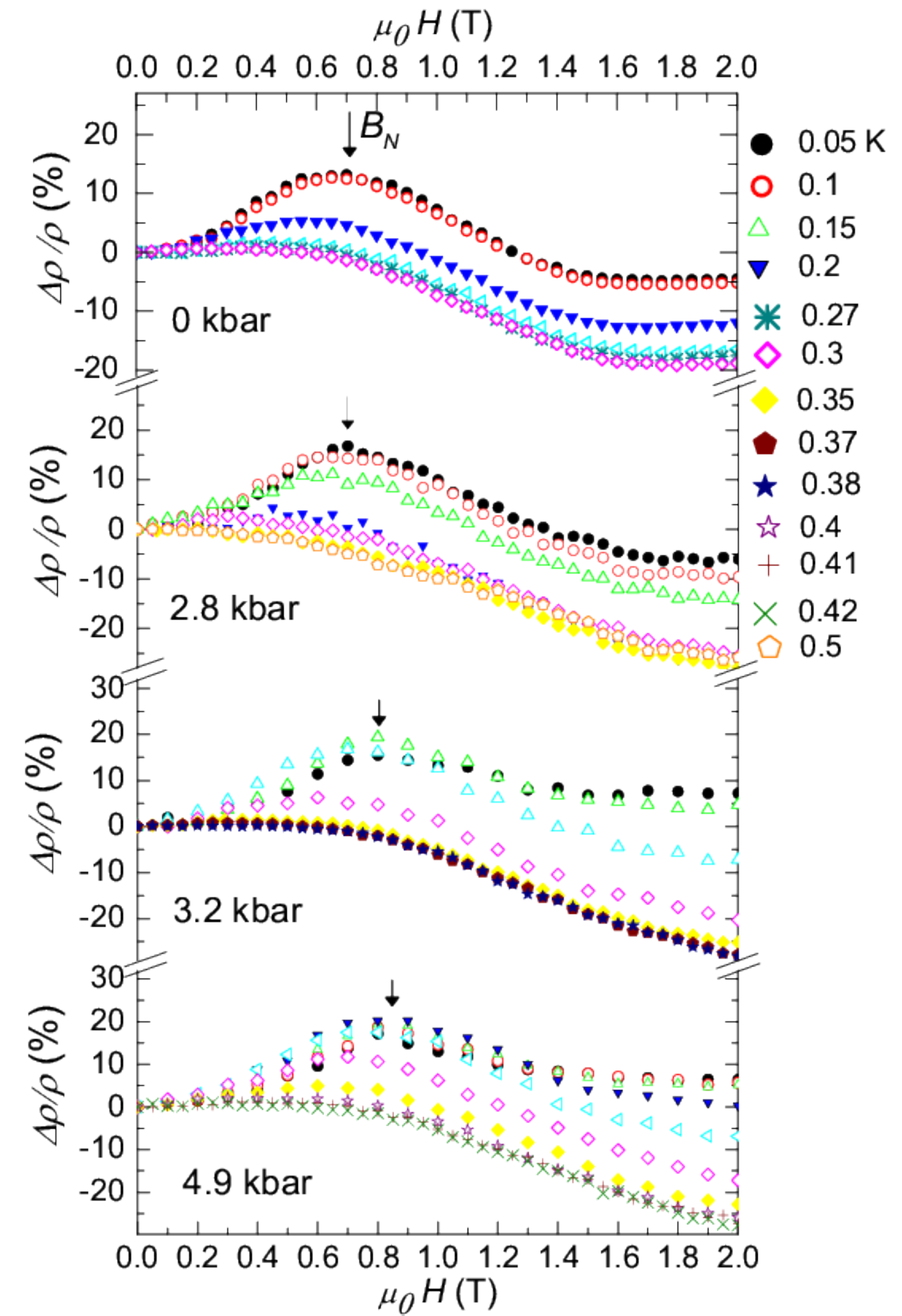}

\caption{(Color online) (a) Isothermal magnetoresistance ($\Delta\rho/\rho$) at selected pressures. The arrows show the magnetic field where $\Delta\rho/\rho$ is maximum ($B_N$) at $T =$ 0.05 K. The value of $B_N$ at different temperatures and pressures is plotted in Fig.~\ref{fig6}. \label{fig5}}

\end{figure}

The relative change of electrical resistivity from $T \rightarrow$ 0 to $T^{\rho}_{N}$, $\Delta\rho_{N} / \rho_{0}= (\rho(T^{\rho}_{N}) - \rho_{0}) / \rho_{0}$, as well as the residual resistivity $\rho_{0}$ (Fig.~\ref{fig2}a) are plotted as a function of $p$ in Fig.~\ref{fig4}b. $\rho_{0}$ was determined by least squares fitting of the data below 0.2 K to $\rho(T) = \rho_{0} + AT^{2}$ (Fig.~\ref{fig3}). As the size of $\Delta\rho_{N}/ \rho_{0}$ is generally considered to be a measure of the strength of the AF order  [\onlinecite{lohn},\onlinecite{hegger},\onlinecite{larrea1},\onlinecite{seo},\onlinecite{Rosch}], its increase with pressure confirms that pressure stabilizes the AF order, at least up to 6.2 kbar.

\begin{figure}[htbp]

%\hspace{-3.435 cm}
\centering \includegraphics[angle=0,scale=0.45]{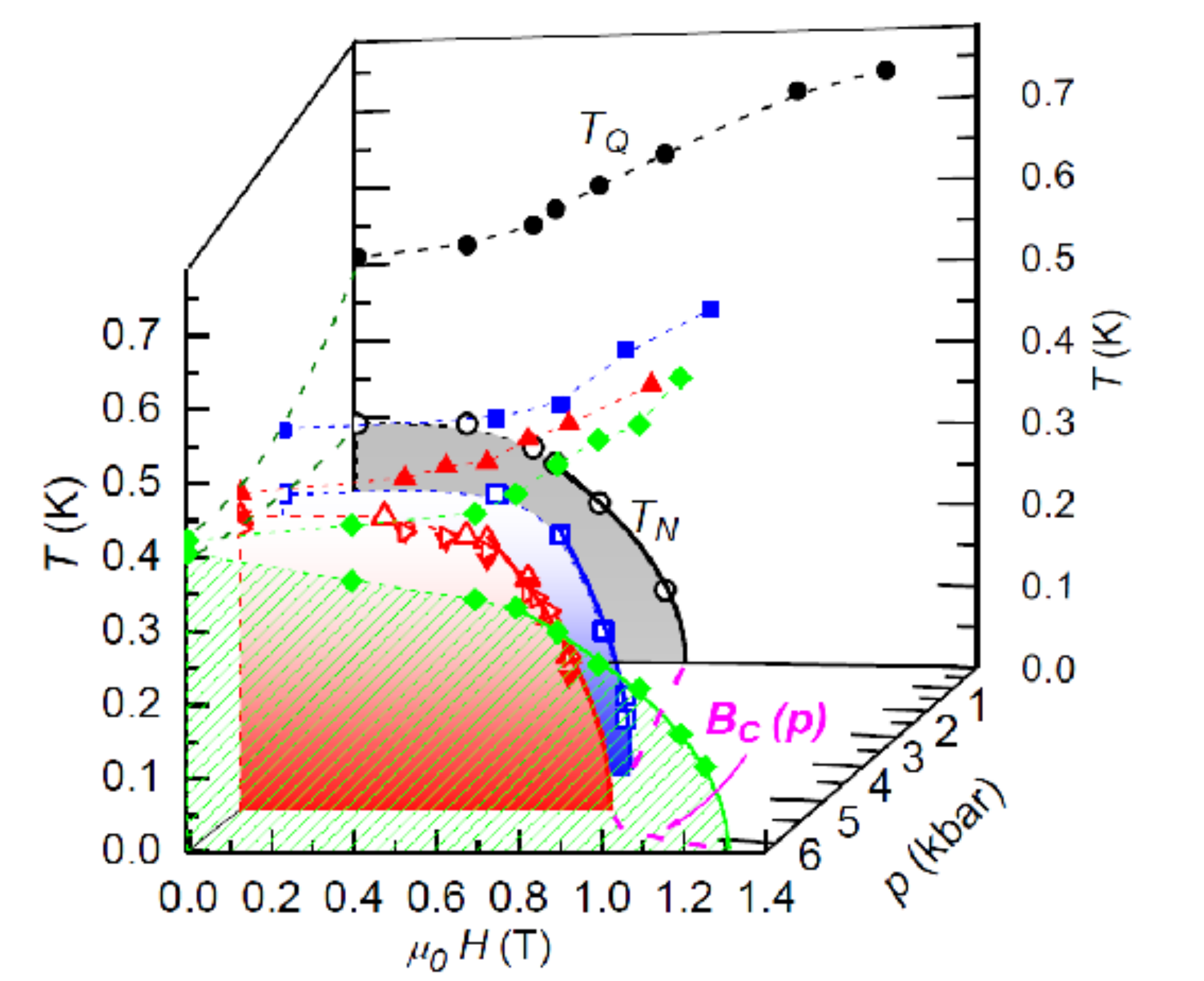}
%\hspace{10.5cm}
\caption{(Color online) Pressure$-$magnetic field$-$temperature phase diagram of Ce$_{3}$Pd$_{20}$Si$_{6}$ determined from our magnetoresistance ($\circ$, $\square$, $\bigtriangleup$), electrical resistivity ($\bigtriangledown$) and specific heat ($\bullet$, $\blacksquare$, $\blacktriangledown$, $\blacktriangle$, $\blacklozenge$ ) measurements. All dashed lines are guide to the eyes. \label{fig6}}

\end{figure}

To determine the pressure$-$magnetic field$-$temperature phase diagram, isothermal magnetoresistance (MR) measurements were done at different pressures (Fig.~\ref{fig5}). For all pressures and for the lowest temperature $T =$ 0.05 K, MR($B$) first increases up to a maximum and then decreases. In Ref. [\onlinecite{cus2012}] $B_N$ was determined by fitting a phenomenological function to MR($B$). Here, for simplicity, we define the position of the maximum as the magnetic field $B_N$ for the suppression of AF order at constant temperature. As temperature increases, $B_N$ is reduced and completely suppressed above $T_N$. $B_N (T)$ isobars define the boundary of the AF phase for different pressures (Fig.~\ref{fig6}).

We can interpret these $B_N (T)$ data also as $T_N (B)$ data and use them to estimate the critical field ($B_{C} = B_{N}$ ($T =$ 0)) with the mean field expression $T_N(B) \propto (B_C - B)^{\frac{1}{2}}$ (full lines in Fig.~\ref{fig6}). $B_C$ increases from 0.75(8) T at $p = $ 0 to about 0.91(2) T at 4.9 kbar. This increase of $B_C$ with pressure supports the strengthening of the AF order with pressure.

\subsection{Specific heat}\label{}

\begin{figure}[htbp]

\centering \includegraphics[angle=0,scale=0.43]{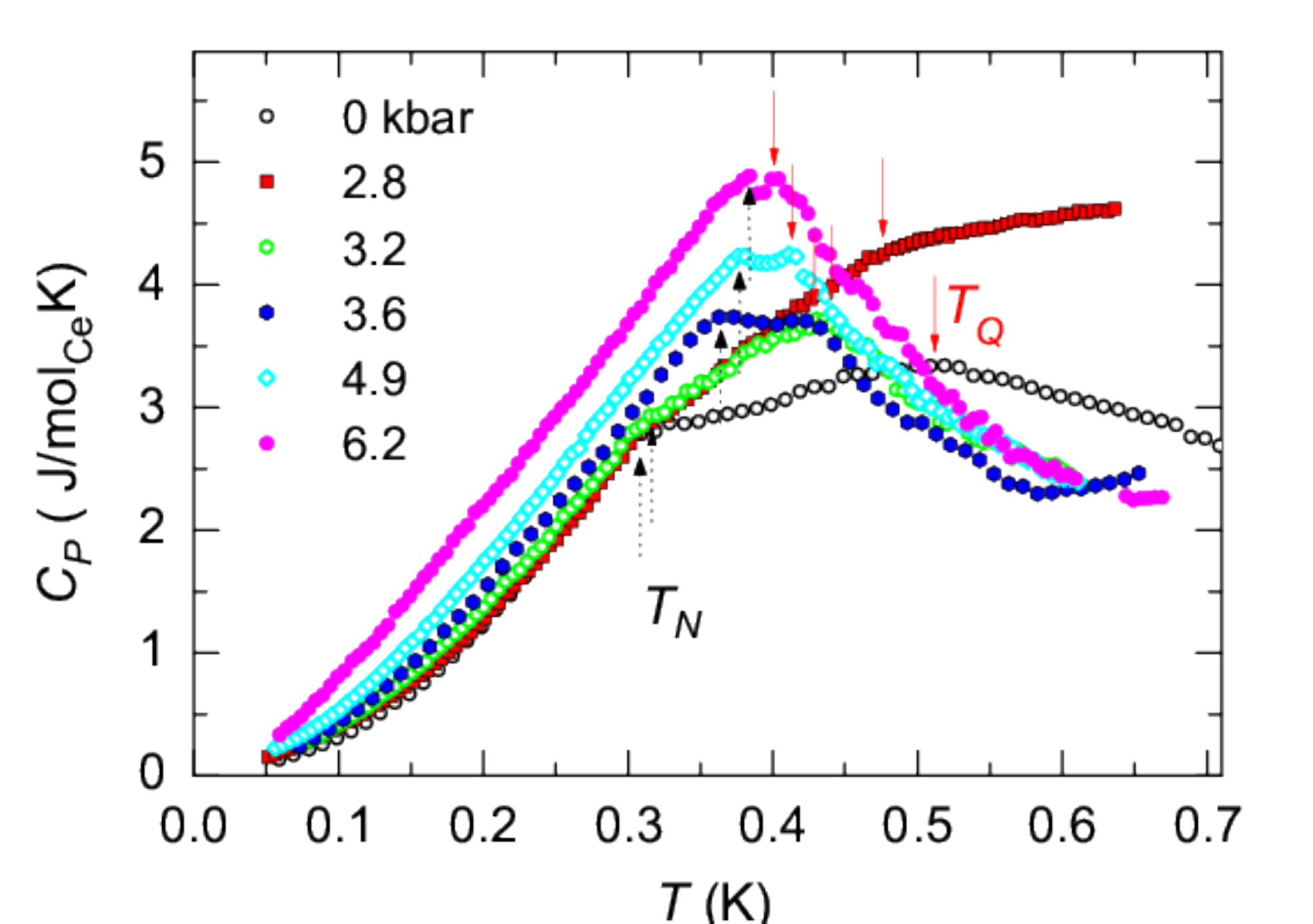}

\caption{(Color online) Temperature variation of the specific heat $C_{p}$ at different pressures. The upward and downward arrows indicate the transitions at $T_{N}$ and $T_{Q}$, respectively. \label{fig7}}

\end{figure}

Our specific heat measurements provide valuable independent information about the evolution of the two phase transitions under pressure. According to the heat dissipation equation, the sample's specific heat ($C_p$) is either proportional to the inverse of the AC-pick up voltage signal ($V_{ac}$) or to the phase shift ($\phi$) measured directly by AC calorimetry [\onlinecite{Wilhelm2003, Gmelin1997,Knebel2009}]. In the pressure range investigated here, the employed thermocouple and heater have negligible pressure variation [\onlinecite{choi2002}] and we can detect the absolute value of phase transition temperatures with an accuracy of 4 $\%$. An independent specific heat measurement was performed at 0.1 Hz to estimate the addenda contribution. The specific heat measured in the pressure cell at $p =$ 0 kbar agrees with the one measured under adiabatic conditions [\onlinecite{prok2009}] (below 0.2 K) if $C_p$ is scaled by a factor 1.5. This calibrates the absolute values of the AC specific heat for all pressures.

\begin{figure}[htbp]

\centering \includegraphics[angle=0,scale=0.4]{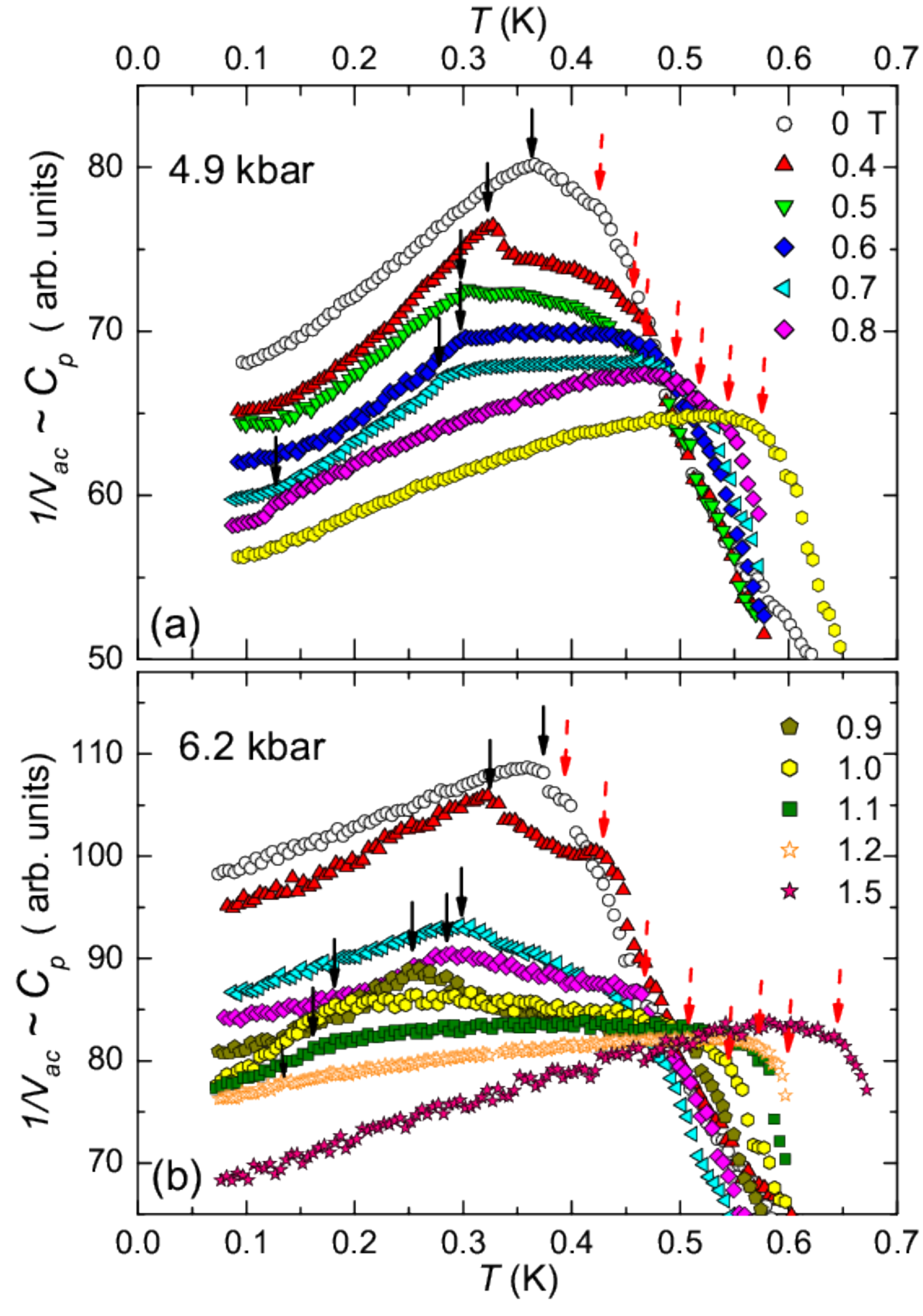}

\caption{(Color online) Temperature dependence of the inverse of the pick-up voltage signal ($V_{ac}$ $\sim$ $C_{p}$), at different magnetic fields and at constant pressures (a) $p =$ 4.9 kbar and (b) $p =$ 6.2 kbar. All these isofield data are shown with the addenda contribution. The solid and dashed arrows show the two observed anomalies at $T_N$ and $T_Q$, respectively (see text). \label{fig8}}

\end{figure}

Figure~\ref{fig7} shows the temperature dependence of the sample's specific heat at different pressures. At $p =$ 0 kbar, two anomalies are observed at $T_{N} =$ 0.3 K and $T_{Q} =$ 0.5 K, associated tentatively with an antiferromagnetic and an antiferroquadrupolar transition, respectively. $T_{N}$ and $T_{Q}$ are determined as the temperatures where $dC_p(T)/dT$ has local maxima (not shown). $T_{N}(p)$ and $T_{Q}(p)$ are plotted in Fig.~\ref{fig4}d. The increase of $p$ has opposite effects on the two anomalies: $T_{N}$ increases whereas $T_{Q}$ decreases. At 6.2 kbar, both transitions seem to merge, being essentially undistinguishable at pressures close to 6.2 kbar.

The application of magnetic field helps to identify both transitions because field is known to suppress $T_{N}$ and to enhance $T_{Q}$ at 0 kbar and low fields [\onlinecite{cus2012}]. Figure~\ref{fig8} shows $1/V_{ac}$ $\sim$ $C_p$ at 4.9 and 6.2 kbar, in different magnetic fields. The frequency of the excitation current for our AC calorimeter was kept constant for all temperature scans. No measurements were done at lower frequencies to estimate the addenda contribution. Thus, Fig.~\ref{fig8} depicts isofield $C_p$ curves without the substraction of an addenda contribution. As we only extract phase transition temperatures from these data this procedure is fully justified. At 4.9 kbar and 0 T (Fig.~\ref{fig8}a), $C_p$ shows two different anomalies at 0.37 K and 0.43 K. The lower anomaly is monotonically shifted to lower temperatures and is completely suppressed at a field slightly above 0.8 T. The upper transition is continuously shifted to higher temperatures. In analogy with the data at $p =$ 0 this identifies the lower and upper transitions as $T_N$ and $T_Q$, respectively. Similar behavior is observed at 6.2 kbar (Fig.~\ref{fig8}b) which confirms that the single broadened anomaly indeed still contains both transitions. $T_N(B)$ and $T_Q(B)$ are plotted in the pressure$-$magnetic field$-$temperature phase diagram in Fig.~\ref{fig6}. $T_N(B)$ obtained from our $C_p$ measurements is in good agreement with the results from our MR measurements. For 6.2 kbar, where no MR data are available, we use the same mean field fit (solid line in Fig.~\ref{fig6}) as above, $T_N(B) \propto (B-B_{C})^{\frac{1}{2}}$ and obtain a critical field $B_{C} =$ 1.32(3) T. Thus, the AF order is strengthened under pressure up to at least 6.2 kbar.

\subsection{Quantum criticality under $B$ and $p$}\label{}

\begin{figure}[htbp]

\centering \includegraphics[angle=0,scale=0.42]{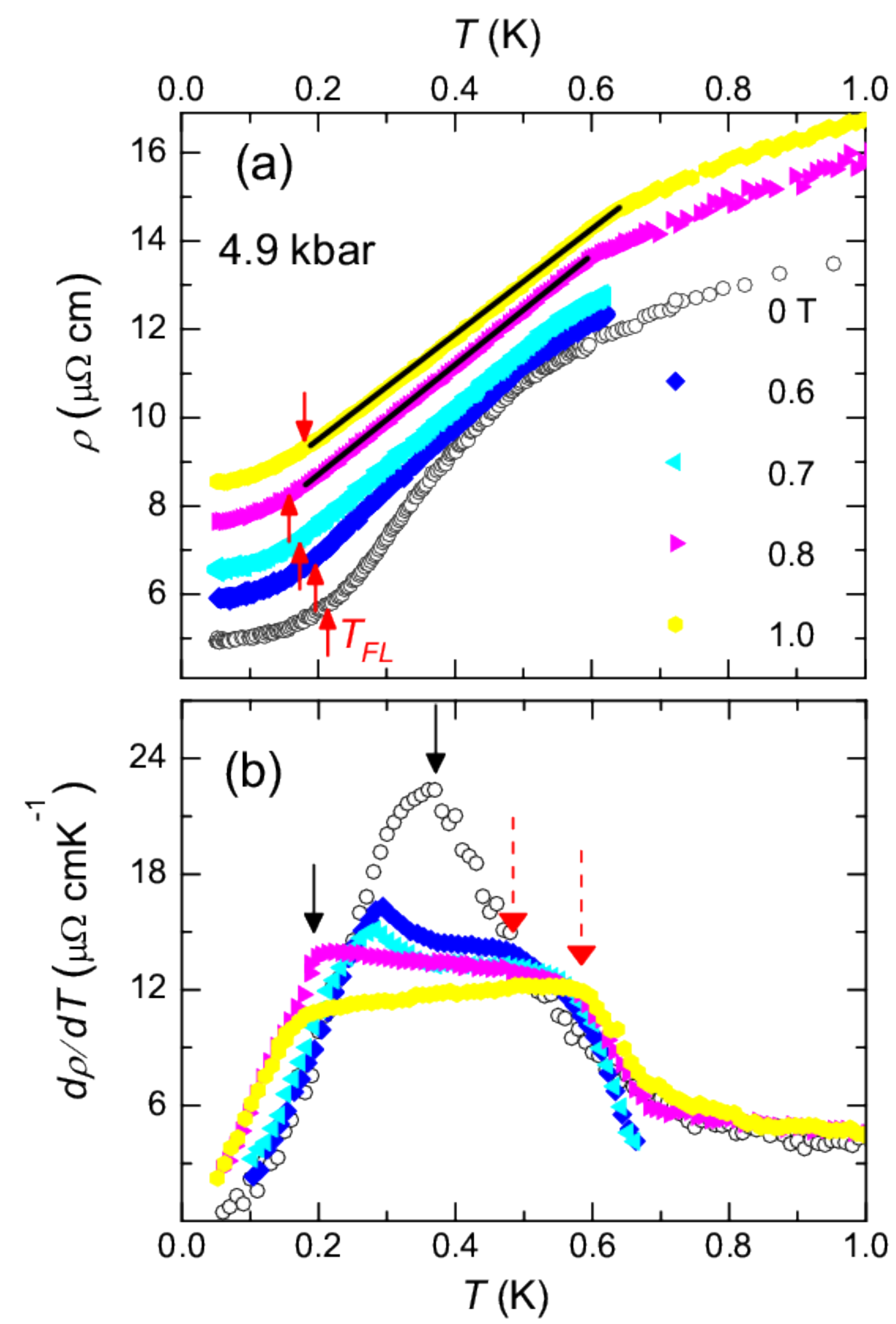}

\caption{(Color online) (a) Temperature variation of the electrical resistivity at different magnetic fields for $p =$ 4.9 kbar. For better readability, $\rho(T)$ data are shifted by +0.5 at 0.6 T, +1 at 0.7 T, +2 at 0.8 T and +3 at 1 T ($\mu \Omega$cm) The straight line indicates the most extended linear temperature range ascribed to non-Fermi liquid behavior. The arrows indicate the Landau-Fermi liquid temperature $T_{FL}$ (see text). (b) Temperature dependence of the first derivative of electrical resistivity ($d\rho/dT$). Solid and dashed arrows indicates putative $T_N$ and $T_Q$, respectively.
\label{fig9}}

\end{figure}

The pressure$-$magnetic field$-$temperature phase diagram (Fig.~\ref{fig6}) hosts a line of zero temperature phase transitions $B_{C}(p)$ from which quantum critical behavior might emerge. This has been shown to be the case for $B_{C}$($p =$ 0) $\simeq$ 0.9 T [\onlinecite{cus2012}]. Here, we investigate the behavior at 4.9 kbar, a pressure where $T_{N}$ and $T_{Q}$ are close to each other (0.37 and 0.41, respectively) in zero magnetic field. Figure~\ref{fig9}a shows the corresponding electrical resistivity data at different magnetic fields. Similar to $\rho(T)$ in Fig.~\ref{fig2}a, an $S$-shaped $\rho(T)$ curve below 0.7 K indicates the presence of AF order below $T_{N}(B)$ from 0 up to 0.8 T. $T_{N}(B)$ is estimated as the temperature where $d\rho(T)/dT$ has a maximum (marked with solid arrows in Fig.~\ref{fig9}b). The $T_{N}(B)$ values are in good agreement with the values extracted above from MR and $C_{p}$ measurements (Fig.~\ref{fig6}). For $B=$ 1 T, the low temperature maximum in $d\rho(T)/dT$ is absent. Thus 1 T is above the critical field for the suppression of the AF order. $T_{Q}$ is seen as a shoulder in $d\rho(T)/dT$ at higher temperatures and finite fields (dashed arrows in Fig.~\ref{fig9}b). It shifts to higher temperatures with increasing field, in agreement with $T_{Q}(B)$ extracted above from $C_{p}$ at 4.9 kbar (Fig.~\ref{fig8}a).

\begin{figure}[htbp]

\centering \includegraphics[angle=0,scale=0.42]{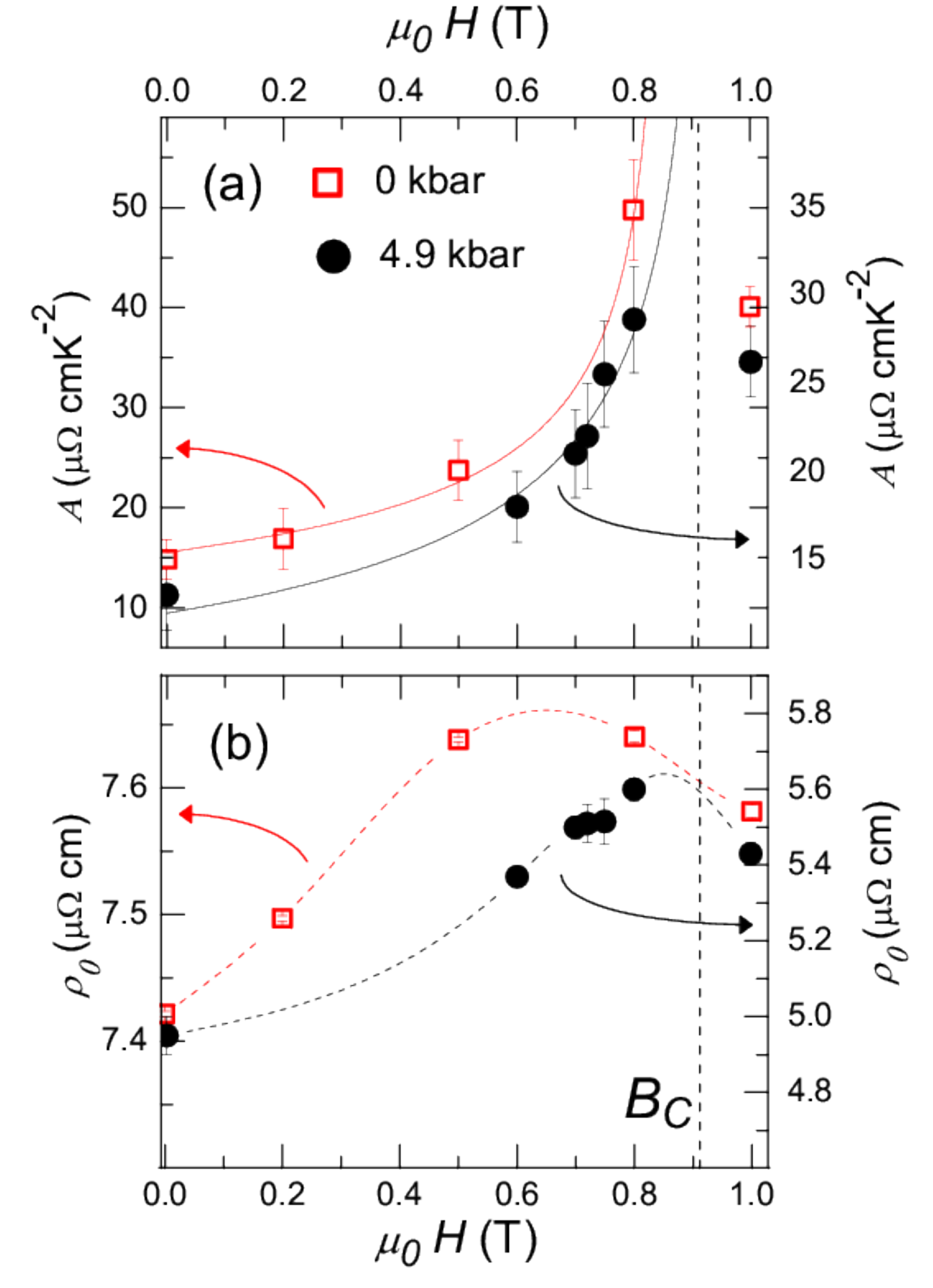}

\caption{(Color online) (a) Landau Fermi liquid $A$ coefficient as a function of magnetic field for 4.9 kbar, and for 0 kbar [\onlinecite{lorenzer2012}] for comparison. The solid curves are fits using the expression $A(B) \propto |B_{C}-B|^{\eta}$ (see text). (b) Residual resistivity $\rho_{0}$ as a function of magnetic field for 4.9 kbar and for 0 kbar. Dashed curves are guide to the eyes whereas straight dashed lines indicate the critical field at $B_{C}$. \label{fig10}}

\end{figure}

To search for signs of field-induced quantum criticality at $p =$ 4.9 kbar we first analyse the Landau-Fermi liquid (LFL) behavior. $\rho(T) = \rho_{0} + A T^{2}$ best explains the lowest temperature electrical resistivity, indicating the absence of significant magnon scattering and the dominance of LFL behavior in the various phases. The $A$ coefficient is determined by a least squares linear fit of $\rho$ plotted vs. $T^{2}$ (not shown) up to the temperature where the fit deviates by more than 0.2 $\%$ from the data. This temperature ($T_{FL}$) is indicated by arrows in Fig.~\ref{fig9}. The values $A$ and $\rho_{0}$ obtained from these fits are plotted in Fig.~\ref{fig10}a,b. For comparison we also show the $A$ and $\rho_{0}$ values of a sample at ambient pressure, which has a slightly higher AF transition temperature ($T_{N} =$ 0.35 K) [\onlinecite{lorenzer2012}].

At $p =$ 4.9 kbar, the strong enhancement of $A$ from 0 to 0.8 T and the smaller $A$ value at 1 T point to a QCP between 0.8 and 1 T. A fit using the expression $A(B) \propto |B_{C}-B|^{\eta}$ [\onlinecite{Moriya}] describes the enhancement of $A(B)$ below 1 T very well (solid line in Fig.~\ref{fig9}a) with $B_{C} =$ 0.91 T and $\eta \approx$ -0.42. A similar value of $\eta$ and $B_{C} =$ 0.87 T was reported for $A(B)$ at ambient pressure (Fig.~\ref{fig10}a) [\onlinecite{lorenzer2012}]. Moreover, at $p =$ 4.9 kbar, the residual resistivity $\rho_{0}$ is only slightly enhanced towards $B_{C}$, with 4.92 $\mu\Omega$cm at 0 T and 5.6 $\mu\Omega$cm at 0.8 T (Fig.~\ref{fig10}b). We observe that the relative enhancement from 0 T up to 0.8 T is $\frac{\Delta \rho_{0}}{\rho_{0}(B=0)}$ = 0.03 and $\frac{\Delta A}{A (B=0)}$ = 2.3 for 0 kbar, and $\frac{\Delta \rho_{0}}{\rho_{0}(B=0)}$ = 0.13 and $\frac{\Delta A}{A (B=0)}$ = 1.2 for 4.9 kbar.

Finally, we analyse deviations from the LFL behavior. $\rho(T)$ at 4.9 kbar and $B =$ 1.0 T is linear in temperature from below 0.18 K to 0.6 K. The range of linear $T$-dependence slightly shrinks at 0.8 T where it persists from below 0.2 K up to 0.56 K. The non-Fermi liquid temperature dependence $\rho \sim T$ has frequently been observed in systems with Kondo breakdown QCP [\onlinecite{lohn},\onlinecite{paschen2007}]. The latter together with the enhancement of $A(B)$ and $\rho_{0}(B)$ towards $B_{C} =$ 0.91 T indicates a field-induced QCP for $p =$ 4.9 kbar, that is similar to one seen at ambient pressure.

\subsection{Temperature-pressure phase diagram}\label{}

The temperature$-$pressure ($T-p$) phase diagram in Fig.~\ref{fig4}d shows the converse effects of pressure of $T_{N}$ and $T_{Q}$: $T_{N}$ increases with pressure ($\Delta T_{N}(p) = T_{N}(p =$ 6.2 kbar) $- T_{N}(p = 0) \approx$ +0.1 K) whereas $T_{Q}$ decreases with pressure ($\Delta T_{Q}(p) \approx$ -0.1 K). This phase diagram is distinctly different from the $T-p$ phase diagrams reported for the related compounds CeB$_{6}$ [\onlinecite{brandt1985}] and Ce$_{3}$Pd$_{20}$Ge$_{6}$ [\onlinecite{hidaka}]. These are cubic heavy fermion compounds that were shown to undergo magnetic and quadrupolar transitions [\onlinecite{paschen2014}]. In CeB$_{6}$, $T_{N}$ decreases whereas $T_{Q}$ increases with pressure up to 10 kbar [\onlinecite{brandt1985}], which is a trend opposite to what we observe for Ce$_{3}$Pd$_{20}$Si$_{6}$. For Ce$_{3}$Pd$_{20}$Ge$_{6}$, $T_{N}$ and $T_{Q}$ at first increase and subsequently decrease with pressure [\onlinecite{hidaka}]. In spite of the similarities of the zero-pressure$-$temperature$-$field phase diagrams of all these compounds, this points to different origins thereof.

\begin{figure}[htbp]
\hspace{-1.5 cm}
\centering \includegraphics[angle=0,scale=0.5]{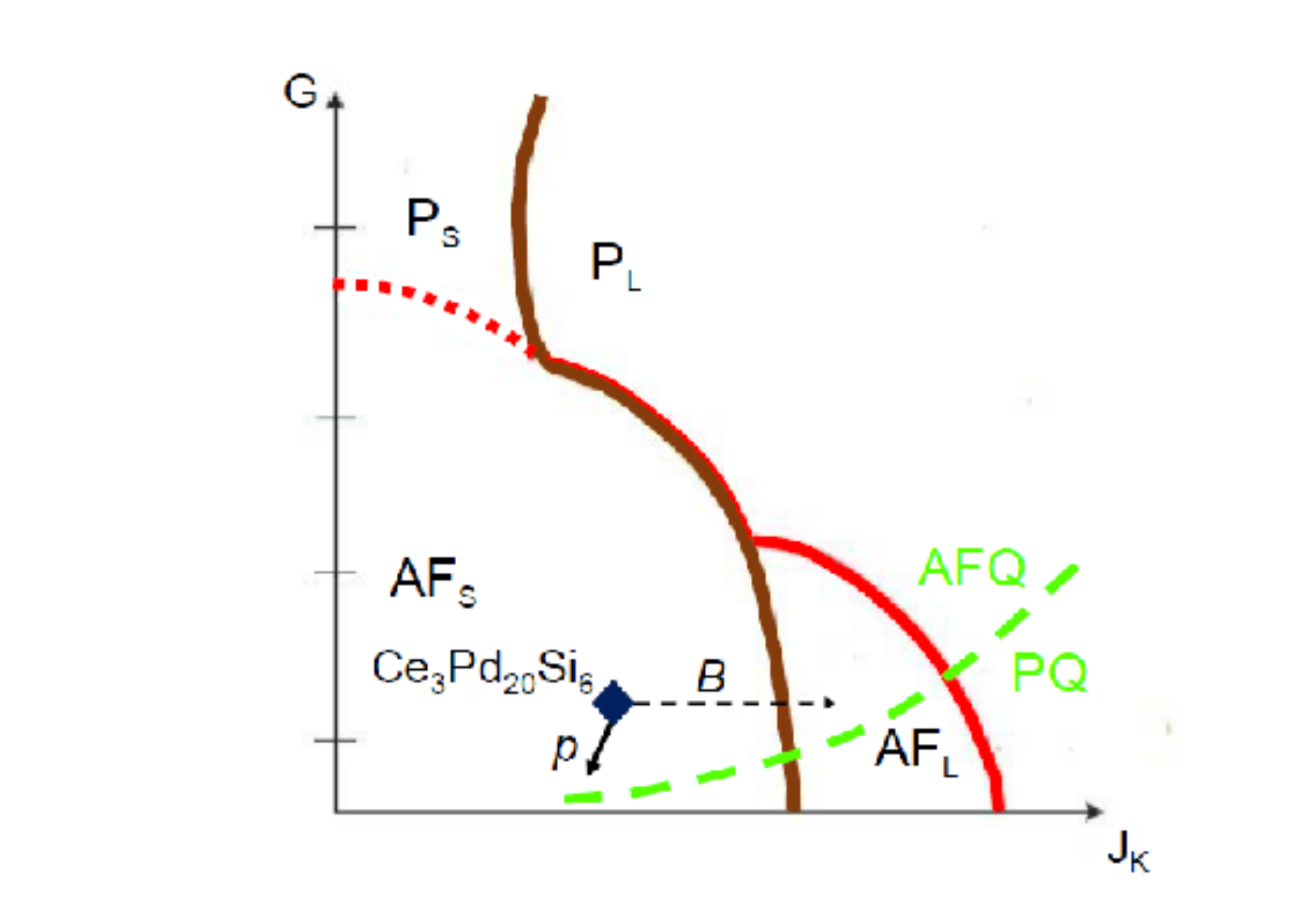}

\caption{(Color online) Global $T =$ 0 phase diagram (GPD) for heavy fermion compounds close to an antiferromagnetic instability spanned by the magnetic frustration parameter $G$ and the Kondo coupling $J_{K}$ [\onlinecite{si2010},\onlinecite{cus2012}]. Lines of quantum critical points separate antiferromagnetic (AF) from paramagnetic (P) (thick red lines), and regions of small (S) and large (L) Fermi surfaces (brown line). The latter represents quantum critical points accompained by Kondo breakdown. The diamond solid symbol represents the position of Ce$_{3}$Pd$_{20}$Si$_{6}$ in the GPD at ambient conditions ($p =$ $B$ = 0). The solid and dashed arrows represent the direction that Ce$_{3}$Pd$_{20}$Si$_{6}$ follows under pressure and magnetic field, respectively. The thick dashed line tentatively locates the boundary between a phase with (AFQ) and a phase without (PQ) antiferroquadrupolar order. \label{fig11}}

\end{figure}

We now try to rationalize our findings within the framework of the global phase diagram for antiferromagnetic heavy fermion compounds [\onlinecite{si2010},\onlinecite{cus2012}]. This is a $T =$ 0 two-dimensional phase diagram spanned by the magnetic frustration parameter $G$ and the Kondo coupling constant $J_K$. In systems with antiferromagnetic order, the antiferromagnetic phase expands with decreasing $G$. It has been previously shown that the ground state of Ce$_{3}$Pd$_{20}$Si$_{6}$ at $p =$ 0 and $B =$ 0 is located in the AF ordered regime with small Fermi surface (AF$_S$). At $p =$ 0, $B$ drives Ce$_{3}$Pd$_{20}$Si$_{6}$ to another ordered (presumably still AF) state with a large Fermi surface (AF$_L$), passing through a Kondo breakdown QCP. This AF-QCP has been associated with an increase of $J_K$ at constant $G$ [\onlinecite{cus2012}]. Our experiments revealed that the field to reach this QCP is increased with pressure. This suggests that pressure drives Ce$_{3}$Pd$_{20}$Si$_{6}$ even deeper into the antiferromagnetic phase (Fig.~\ref{fig11}). Simultaneous application of pressure and field induces a quantum critical point with quantum critical resistivity characteristics very similar to the $p =$ 0 case. This indicates that the criticality remains dominated by $T_{N}$ even though $T_{Q}$ is sizably decreased by pressure. The quadrupolar phase boundary is not captured by the present version of the theoretical global phase diagram. Our experiments reveal that its dependence on $G$ and $J_K$ is very different from the phase boundary between AF$_S$ and AF$_L$. To visualize this, we tentatively draw a line of quantum critical points between an AFQ phase and a phase without quadrupolar order (PQ, Fig.~\ref{fig11}). It shows that pressure tuning ultimately exposes a quadrupolar QCP within an AF background, which is an exciting prospect for future research.

 \section{Conclusions}\label{}

To summarize, we have investigated the pressure evolution of the putative antiferromagnetic and antiferro-quadrupolar orders in Ce$_{3}$Pd$_{20}$Si$_{6}$ using electrical resistivity, magnetoresistance and specific heat measurements. Our results reveal an increase of the antiferromagnetic ($T_{N}$) and a decrease of the antiferroquadroplar ($T_{Q}$) ordering temperatures with pressure and the merging of both transitions at about 6.2 kbar. This converse effect of pressure on $T_{N}$ and $T_{Q}$ is rather unique in cubic heavy fermion compounds.

At pressures where $T_{N}$ $\approx$ $T_{Q}$, the application of magnetic field induces a QCP, with a critical field $B_{C}$ that is larger than at $p =$ 0 but with the same quantum critical $\rho(T)$ behavior. Our findings are consistent with pressure moving the location of Ce$_{3}$Pd$_{20}$Si$_{6}$ in the global phase diagram for quantum critical heavy fermion compounds towards lower values of the frustration parameter $G$ and the Kondo coupling $J_{K}$. This would imply that the role of pressure is to enhance the three-dimensional character of the low-lying magnetic and quadrupolar interactions.

Finally, our experimental findings qualify pressure as ideal tool to disentangle effects of dipolar and higher multipolar ordering, and quantum criticality emerging from their suppression. This will likely trigger further experiments in higher pressures.

%\acknowledgments{We acknowledge the European Research Council (ERC Advanced Grant No 227378) for financial support. J. Larrea J. acknowledges the FRC/URC of UJ for funding of a Postdoctoral Fellowship under joint supervision of SP and AMS. JLJ also acknowledges the CNPq/MCTI-Brazil and V. Martelli acknowledges FAPERJ (Nota 10).}

\acknowledgments{We acknowledge the European Research Council (ERC Advanced Grant No 227378) and the Austrian Science Fund (FWF Grant I623-N16) for financial support. JLJ acknowledges the FRC/URC of UJ for funding of a Postdoctoral Fellowship under joint supervision of SP and AMS. JLJ also acknowledges the CNPq/MCTI-Brazil and VM acknowledges FAPERJ (Nota 10).}

\bibliography{bib_JLJ}% Produces the bibliography via BibTeX.]
\bibliographystyle{apsrev4-1}

\end{document}